\documentclass[sigconf]{acmart}
\AtBeginDocument{%
  \providecommand\BibTeX{{%
    \normalfont B\kern-0.5em{\scshape i\kern-0.25em b}\kern-0.8em\TeX}}}

\copyrightyear{2022} 
\acmYear{2022} 
\setcopyright{acmlicensed}\acmConference[SA '22 Technical Communications]{SIGGRAPH Asia 2022 Technical Communications}{December 6--9, 2022}{Daegu, Republic of Korea}
\acmBooktitle{SIGGRAPH Asia 2022 Technical Communications (SA '22 Technical Communications), December 6--9, 2022, Daegu, Republic of Korea}
\acmPrice{15.00}
\acmDOI{10.1145/3550340.3564218}
\acmISBN{978-1-4503-9465-9/22/12}

\setcitestyle{square}

\begin{document}

\title{Rig Inversion by Training a Differentiable Rig Function}

\author{Mathieu Marquis Bolduc}
\email{mmarquisbolduc@ea.com}
\affiliation{%
  \institution{SEED Electronic Arts}
 % \streetaddress{P.O. Box 1212}
  \city{Austin}
  \state{Texas}
  \country{United States of America}
}

\author{Hau Nghiep Phan}
\email{hphan@ea.com}
\affiliation{%
  \institution{SEED Electronic Arts}
 % \streetaddress{P.O. Box 1212}
  \city{Montr\'eal}
  \state{Qu\'ebec}
  \country{Canada}
}

\renewcommand{\shortauthors}{Marquis Bolduc, et al.}

\begin{abstract}
  Rig inversion is the problem of creating a method that can find the rig parameter vector that best approximates a given input mesh. In this paper we propose to solve this problem by first obtaining a differentiable rig function by training a multi layer perceptron to approximate the rig function. This differentiable rig function can then be used to train a deep learning model of rig inversion.
\end{abstract}

\begin{CCSXML}
<ccs2012>
   <concept>
       <concept_id>10010147.10010257</concept_id>
       <concept_desc>Computing methodologies~Machine learning</concept_desc>
       <concept_significance>500</concept_significance>
       </concept>
   <concept>
       <concept_id>10010147.10010371.10010352</concept_id>
       <concept_desc>Computing methodologies~Animation</concept_desc>
       <concept_significance>500</concept_significance>
       </concept>
 </ccs2012>
\end{CCSXML}

\ccsdesc[500]{Computing methodologies~Machine learning}
\ccsdesc[500]{Computing methodologies~Animation}

\keywords{rig inversion, neural networks, computer animation}


\maketitle

\section{Introduction}
It is possible to directly obtain 3D and 4D mesh data from various capture systems (e.g.\cite{capture, capturestatic}) as an alternative to traditional motion capture that tracks joints positions. One motivation for such systems is to capture soft body deformations at a high fidelity. It is desirable for this 3D data to be rigged, i.e. to obtain the set of best-corresponding animation parameters (such as animation tool GUI parameters). This allows 3D animation artists to edit the pose using their own tools, as well as to use the captured data in various tools and software built for any particular rig. If the rig logic is a function that outputs a 3D mesh from animation parameters inputs \cite{rigspace}, or rig parameters, then obtaining the rig parameters corresponding to a 3D mesh is the problem of inverting that function. It should be noted that the rig logic that we wish to inverse may be neither injective nor surjective; it may be possible for the exact same mesh to be obtained from different sets of rig parameters, and it is also extremely unlikely for the rig function to be able to perfectly output any given mesh that can be captured. The consequence is that in practice the rig function should not be considered bijective, i.e. it cannot be perfectly inverted. Thus the problem of rig inversion becomes the problem of finding a function whose output are the rig parameters that, as input to the rig logic, minimize the difference between the input mesh and the rigged mesh.

\section{Previous Work}
Rig inversion has been a subject of interest in recent years. Noting that the performance costs of Jacobian-based methods grows to the square of the number of training samples, \cite{holden1} proposed a more memory-efficient method based on training a multi-layer perceptron to minimize the difference in output rig parameters given an input mesh, using data from animators. \cite{holden2} also proposed a method based on Gaussian process regression. \cite{clustering} clusters vertices together with corresponding rig control parameters such that each cluster can be inverted using gaussian processes regression. \cite{rackovic2} improves on linear approximations of the rig by modeling the rig using a second-order polynomial which is then inverted using Levenberg-Marquardt optimization. \cite{gustafson} analytically create an inference-optimized version of their rig as to speed up jacobian-based methods of rig inversion.

\section{Proposed Method}
\label{sec:proposemethod}

In this work we assume that the rig logic \(L(r)\) (Eq. \ref{equation:one}), which returns the mesh \(x\) from rig parameters values \(r\), is complex and non-linear. This is the case for the facial animation rigs we use, and thus the problem of rig inversion cannot be solved by linear methods. We also assume rigs that have too many parameters to be solved in a practical manner by Jacobian-based methods. We note that our rigs cannot be completely clustered as proposed in \cite{clustering} because they don't respect that work’s clustering hypothesis. When we attempted to invert our rigs with meshes from capture sessions using previously published methods such as \cite{holden1}, we were faced with a number of difficulties, which we will now detail.

\begin{equation}
\label{equation:one}
L(r) = x
\end{equation}
\begin{equation}
\label{equation:two}
L^{-1}(L(r)) = r
\end{equation}
\begin{equation}
\label{equation:three}
L(\hat{L}^{-1}(x)) \approx x
\end{equation}

\subsection{Issues in inverting the rig function}

\subsubsection{Non-Surjectivity}

The limited data from rigged animations does not always generalize well to new poses that are not observed during training. Even with a comprehensive training set, because our rig functions are not surjective, there are captured meshes that do not exist in the space of rigged meshed, and thus cannot be represented in a training dataset built from rigged animations. As discussed in section \ref{sec:proposemethod}, in these cases where an exact solution \(L^{-1}\) does not exist (Eq. \ref{equation:two}), we wish to instead find an approximate inverse rig function \(\hat{L}^{-1}(x)\) that will provide the rig parameter vector \(r\) that best approximates the input mesh \(x\) when applied to the rig logic \(L\) (Eq. \ref{equation:three})

\subsubsection{Non-Injectivity}

Our rig functions are not injective, i.e. there may exist a set of different input rig parameters that will output the same mesh. This means that training a deep learning model to invert such rigs using a regression loss on the rig parameters is noisy, as the training dataset will contain different ground truths for identical or very similar inputs. In practice this means gradients from that loss will make the model learn to output an average of such ground truths, with no guarantees that these averages will be good solutions.

\subsubsection{Distribution of rig parameters}
\label{sec:distribution}

In practice, not every value in the rig parameter vector \(r\) has the same importance. Some rig parameters have a more pronounced effect on the resulting mesh than others. While it is desirable to learn all of them, some can tolerate a higher error rate when we wish to minimize the perceptual difference between a mesh \(x\) and \(L(\hat{L}^{-1}(x))\). However, the importance of each parameter is difficult to weigh and prone to human bias.

\subsection{Obtaining a differentiable rig function}

Our proposed method is based on the idea that all three of these difficulties can be solved by the same change to the training procedure. We propose to train the deep learning model inverting the rig function by replacing the loss on the output rig parameter vector \(r\) by a loss on the mesh \(L(\hat{L}^{-1}(x))\) resulting from the rig parameter vector \(L^{-1}\) when evaluated by the rig function \(L\). This can be made practical by obtaining a differentiable approximation \(\hat{L}_d\) of the rig function \(L\). This approximation can then be used after the inverted rig model during training in an Encoder-Decoder fashion (Fig. \ref{fig:inversion}) to obtain an output mesh. A loss can then be applied on this resulting mesh and back propagated to the inverted rig model.

\begin{figure}
     \centering
     \includegraphics[width=\linewidth]{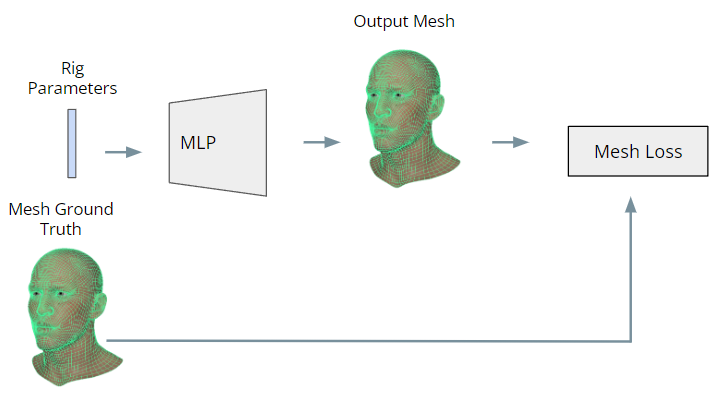}
     \caption{Learning an approximation of the rig function.}
     \Description{The inputs are the rig parameters and a mesh ground truth. An MLP takes the rig parameters as input and outputs a mesh. A loss is applied to the resulting mesh using the mesh ground truth.}
     \label{fig:forward}
\end{figure}

The rig function is most often treated as a black box \cite{holden2, tuning} that can be evaluated but not analytically manipulated. It may also be that the rig function \(L\) cannot be derived for the entire input domain \(\bar{r}\). We can however train a fully differentiable approximation \(\hat{L}_d\) of the rig function. We propose to train a decoder-shaped MLP using a dataset of randomly selected rig parameter vectors that have been evaluated by the rig logic in the animation software (Fig. \ref{fig:forward}). Unlike its inverse \(L^{-1}\), the rig logic \(L\) is assumed to be a true function, and thus training a deep learning model to approximate it is likely to pose less difficulties than approximating its inverse. Recent work provides more details on how to best train an approximation of the rig function \cite{Song_2020}.

\subsection{Training the inverse rig model}

\begin{figure}
     \centering
     \includegraphics[width=\linewidth]{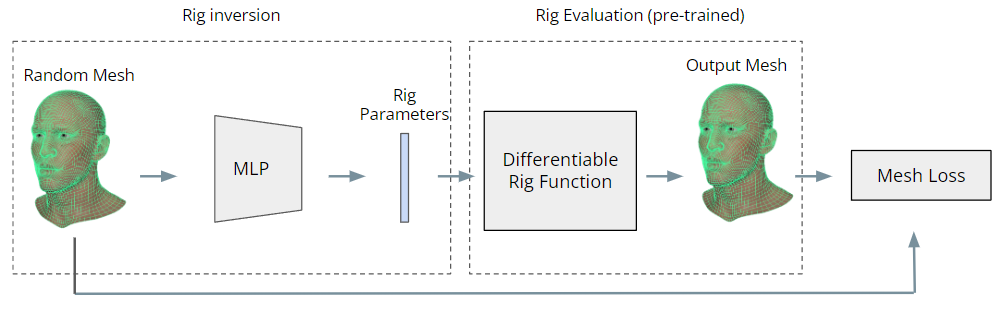}
     \caption{Training the rig inversion model using a differentiable rig approximation.}
     \Description{The input is a random mesh. It is fed to an MLP, which outputs rig parameters. The resulting rig parameters are fed to the differentiable rig function, which outputs a mesh. A mesh loss is applied to this mesh, respective to the input one.}
     \label{fig:inversion}
\end{figure}

Having a differentiable approximation of the rig, we can train the rig inversion model using only mesh data with no corresponding rig parameter vector ground truth (Fig. \ref{fig:inversion}), which solves the three issues we presented at the beginning of this section. To obtain a good generalization, this training data should be randomly augmented (Section \ref{sec:distribution}) to provide noisy but recognizable meshes in the largest possible variety of plausible poses. Regarding the last issue we presented (Section \ref{sec:distribution}), we assume that mesh topologies will be oversampled in regard to the number of vertices in the most perceptually critical areas. The fortunate consequence of this hypothesis is that errors on the most critical rig parameters will result in higher losses on the mesh.

There are a few additional properties to consider. As \cite{holden1} observes, it is desirable to obtain rig parameters that the artists would find sensible. As such, sparse rig parameter are a desirable property as artists will want to minimize the number of parameters to work with. Using Relu activations in the rig inverse model is a good way to encourage sparsity \cite{Goodfellow-et-al-2016}. Finally, it is also desirable to artists for subsequent frames in an animation to have temporarily coherent rig parameters, something that is not guaranteed if the rig function is not bijective. To this end, we may need to enforce a local Lipschitz continuity in the inverse rig function network. Noise augmentation is an effective way of doing so \cite{noise}.

\section{Experimental Results}

\begin{figure}
    \includegraphics[width=\linewidth]{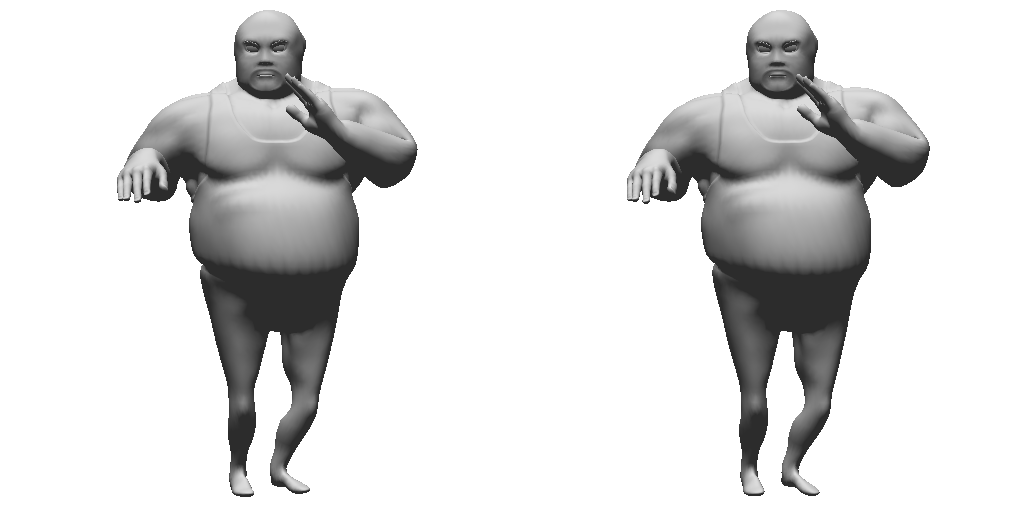}
     \caption{Qualitative result on a full body, non-blendshape rig. The difference between the original animation (left) and the mesh animated with recovered rig parameters (right) is almost imperceptible.}
     \label{fig:body}
\end{figure}

All of the quantitative experimental results we present are from a facial animation rig used by a content creation team at EA. The rig in question has 137 input parameters and outputs 628 weights, which are linearly composed to create a 8447 vertices mesh in a matrix multiplication operation. Additional, qualitative tests using animations instead of capture data confirmed the proposed method is applicable to other kinds of animation rigs as well (Fig. \ref{fig:body}).

\subsection{Differentiable rig function}

From the experimental rig we generated a set of 100000 random rig parameter values. As stated in Section \ref{sec:proposemethod}, sparsity is desirable, and so we select both the number of non-zero values as well as their respective values following a uniform distribution. Those random rig parameter activations are run in Maya™’s python scripting system in order to obtain the corresponding  weights and thus the mesh output. 10\% of this dataset is reserved as a test set and the remaining 90\% as a training set. Previous work such as \cite{holden2} have stated that the size of such a training dataset needs to grow in an impossibly large fashion. This is true only if all inputs and outputs are dependent on every other one. In practice this is rarely the case and not the case with the rig used in these experiments, and as such it is possible for the dataset to have a size that is practical. As the dataset is randomly generated in a uniform distribution, no validation set is used during training.

The architecture used to model \(\hat{L}_d\) is a MLP with two hidden layers of 1024 parameters which outputs 628  factors. Hidden layers have leaky ReLu activations \cite{leaky} while the output has sigmoid activations to constrain the output between [0,1] which is the  weights range. We use no normalization as it severely degrades our experimental results. The model is trained by an Adam optimizer with a learning rate starting at 1e-4 and being halved every time the training loss reaches a plateau for 20 epochs, until the learning rate is smaller than 1e-6. Blendshape factors are multiplied by the  matrix, and the loss used is a MSE loss on the vertex positions of the resulting mesh. All results are consistent across different random weight initialisations.

As the rig we used for experiments is uncharacteristically not a black box, we also compare results with a differentiable implementation of that rig function in Pytorch™.

\begin{table}[ht]
 \caption{Rig logic approximation results presenting vertex error in mm.}
  \centering
  \begin{tabular}{lll}
    \toprule
   % \multicolumn{2}{c}{Part}                   \\
    %\cmidrule(r){1-2}
    Model     & Mean Vertex Error & Maximum Vertex Error \\
    \midrule
    Programmatic & \textbf{0.22} & \textbf{1.4}     \\
    MLP Model     & 0.78 & 5.9      \\
    \bottomrule
  \end{tabular}
  
  \label{tab:table1}
\end{table}

Table \ref{tab:table1} compares the test result for the trained and programmatic rig logic approximations. The learned approximation is fairly precise with less than a millimeter error on average. The non-zero error in the programmatic method is explained by slight differences in curve evaluation between the Pytorch™ and Maya™ implementations.

\subsection{Inverse rig approximation results}
\label{sec:onlineresults}

Both differentiable rig functions (trained  and programmatic) are used to train an inverse rig function approximation model. In both cases the architecture used is an encoder-shaped MLP with 4 hidden layers of [2048, 1024, 512, 256] parameters with leaky ReLu activations, while the output has tanh activations corresponding to the domain of the rig parameters. It is critical to avoid having the inverse rig model feeding out-of-manifold values to the rig model approximation as this will lead to unpredictable behavior. As with the rig function model we use no normalization layers.

\begin{figure}
    \includegraphics[width=\linewidth]{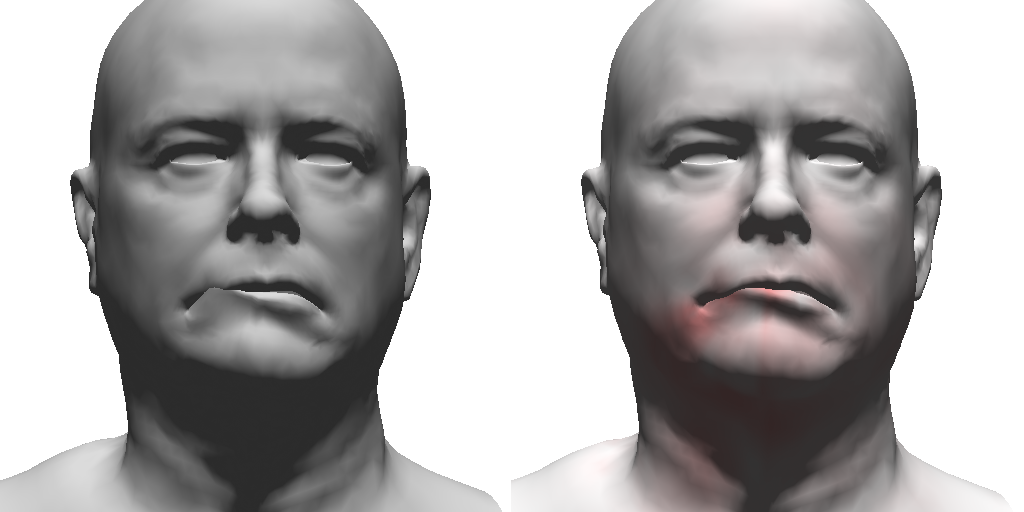}
     \caption{Left: Randomly combining blendshapes with no regard for rig logic produces a recognizable but noisy mesh, similar to imperfect meshes coming from the capture process or content-generation models. Right: Using a mesh loss trains the inverse rig logic to find the rig parameters that best approximate the input mesh when applied to the rig logic, which doubles as a denoising process.}
     \Description{On the left is a noisy head mesh featuring some self-intersecting lips. On the right, putting the mesh on the rig has removed the lip self-intersection.}
     \label{fig:random}
\end{figure}

This model is trained with meshes generated by randomly combining blendshapes with no respect to the rig logic - Since the rig logic is highly non-surjective even in regard to the more limited space of  factors, it is highly unlikely for the rig to be able to exactly represent a random combination of blendshapes (Fig. \ref{fig:random}). Combining all blendshapes usually results in non-plausible meshes, and so both the number of blendshapes used in each training sample as well as their respective value follows a uniform distribution. Each epoch comprises 10000 iterations using mini batches of 64 samples. The model is trained with the same optimizer and schedule as the rig function approximation. All results are consistent across different random weight initialisations.

\begin{table}[ht]
\caption{Rig inversion results presenting vertex error in mm.}
  \centering
  \begin{tabular}{lll}
    \toprule
   % \multicolumn{2}{c}{Part}                   \\
    %\cmidrule(r){1-3}
         & Mean Vertex Error     & Max Vertex Error\\
    \midrule
    \textbf{Random Poses} & & \\
    \midrule
    Programmatic \(\hat{L}_d\) &     \textbf{2.5} & \textbf{15.0}     \\
    Trained \(\hat{L}_d\) & 3.0 & 17.2 \\
    \cite{holden1} & 9.2 & 100.0 \\
    \midrule
    \textbf{Captured Data} & & \\
    Programmatic \(\hat{L}_d\) &     \textbf{4.8} & \textbf{49.0}     \\
    Trained \(\hat{L}_d\) & 4.9 & 49.1 \\
    \cite{holden1} & 10.0 & 99.0 \\

    \bottomrule
\end{tabular}
  
  \label{tab:table2}
\end{table}

Table \ref{tab:table2} compares the result of training the rig inverse approximation using either differentiable rig approximation method and testing the resulting model on a test set of 10000 poses randomly generated using the same method as the training set, as well as a set of captured data. To obtain unbiased test results, we do not use the rig function approximations \(\hat{L}_d\) for testing purposes. Instead, we obtain the rig parameter vector r for each test sample from the inverse rig approximation model and use them as input to the original rig logic \(L\) using the Maya(™) animation software.

Training with either programmatic or trained rig approximations yield a significant improvement over the previous deep learning method of Holden \cite{holden1} which is trained using rig parameters ground truth. Captured data is highly unlikely to be perfectly expressible by the rig, and so achieving a zero error on it is not possible. The capture data is also noisy, which explains the large maximum error for all methods. This is favorable as it allows inverse rig techniques to filter noise in captured data.

\subsection{Parameter Smoothness Results}
\label{sec:smoothnessresults}

Finally we evaluate the usefulness of the model in giving out smooth rig parameters for a temporarily coherent input sequence. We use the squared second-order differences as a measure of roughness, and we compare the result with directly optimizing the rig parameter vector without training an inverse rig model. The results are presented in Table \ref{tab:roughness}. We see that using an encoder-shaped inverse rig model outputs rig parameters that are an order or magnitude smoother, without need for noise augmentation. This is despite each rig parameter vector being initialized with the previous frame value prior to being optimized in order to help with temporal coherency. Qualitatively, the result from directly optimizing the rig parameter vector is too unstable for practical use.

\begin{table}[ht]
  \caption{Temporal coherency results presenting rig parameters roughness}
  \centering
  \begin{tabular}{ll}
    \toprule
   % \multicolumn{2}{c}{Part}                   \\
    %\cmidrule(r){1-2}
         & Roughness \\
    \midrule
    
    Proposed method with Programmatic \(\hat{L}_d\) &     3.8e-3     \\
    Proposed method with trained \(\hat{L}_d\) & \textbf{3.4e-3} \\
    Directly optimizing with Programmatic \(\hat{L}_d\) &     2.3e-2     \\
    Directly optimizing with trained \(\hat{L}_d\) & 2.7e-2 \\

    \bottomrule
\end{tabular}
 
  \label{tab:roughness}
\end{table}

\section{Conclusion and Future Work}
We demonstrated the effectiveness of using a differentiable rig function approximation to train a rig function approximation. Our proposed method presents several improvements over the prior art. First and foremost, the proposed method fully addresses the non-bijective nature of complex rig functions while also generalizing well to plausible meshes that cannot be exactly produced by the rig function. The proposed method also requires no dataset of  artist-produced rigged poses or animations. As future work we hope to improve accuracy and reduce the number of training epochs by using architectures that are tailored for mesh data instead of MLPs.

\begin{acks}
We want to acknowledge Mattias Teye from SEED Electronic Arts for his valuable contribution to this project. We also want to thank Adobe™ for the luchador character.
\end{acks}

\bibliographystyle{ACM-Reference-Format}
\bibliography{sample-base}

\end{document}